# Deep Learning for Real-Time Aerodynamic Evaluations of Arbitrary Vehicle Shapes

## Sam Jacob Jacob[1], Markus Mrosek[2], Carsten Othmer[2] and Harald Köstler[1]

[1]Friedrich-Alexander University of Erlangen-Nuremberg, Germany

[2]Volkswagen AG, Germany

## Abstract

The aerodynamic optimization process of cars requires multiple iterations between aerodynamicists and stylists. Response Surface Modeling and Reduced-Order Modeling (ROM) are commonly used to eliminate the overhead due to Computational Fluid Dynamics (CFD), leading to faster iterations. However, a primary drawback of these models is that they can work only on the parametrized geometric features they were trained with. This study evaluates if deep learning models can predict the drag coefficient ($c_d$) for an arbitrary input geometry without explicit parameterization. We use two similar data sets (total of 1000 simulations) based on the publicly available DrivAer geometry for training. We use a modified U-Net architecture that uses Signed Distance Fields (SDF) to represent the input geometries. Our models outperform the existing models by at least 11% in prediction accuracy for the drag coefficient. We achieved this improvement by combining multiple data sets that were created using different parameterizations, which is not possible with the methods currently used. We have also shown that it is possible to predict velocity fields and drag coefficient concurrently and that simple data augmentation methods can improve the results. In addition, we have performed an occlusion sensitivity study on our models to understand what information is used to make predictions. From the occlusion sensitivity study, we showed that the models were able to identify the geometric features and have discovered that the model has learned to exploit the global information present in the SDF. In contrast to the currently operational method, where design changes are restricted to the initially defined parameters, this study brings data-driven surrogate models one step closer to the long-term goal of having a model that can be used for approximate aerodynamic evaluation of unseen, arbitrary vehicle shapes, thereby providing complete design freedom to the vehicle stylists.

## Introduction

Simulations have witnessed widespread adoption for aerodynamic optimization processes, with one of the principal goals being to minimize the carbon footprint. Any improvement in the aerodynamic coefficients of a car results in increased fuel savings over its lifetime. It is common to simulate geometries with minor variations for iterative improvement based on prior experience or by trial & error. However, evaluating a few handcrafted simulations might lead to suboptimal designs, and searching large design spaces is computationally expensive.

During the early development stages, approximate aerodynamic results often suffice. Data-driven surrogate models trained using high-fidelity simulations can be used to get fast approximate results in a constrained design space. High-fidelity simulations did not lead to abandoning wind tunnel experiments but increased the number of geometries evaluated. Similarly, data-driven surrogate models can be used as a precursor to high fidelity simulation, enabling the stylists and aerodynamicists to search faster in a broader design space.

For the development of the vehicle shape, not only the aerodynamicist's advice but also the advice regarding aesthetics from the stylist have to be considered. Multiple iterations between stylists & aerodynamicists are typically required for the aerodynamic optimization process. However, the improvements are slow because of the overhead due to the CFD computation time of up to several days. Methods like adjoint sensitivity [1] and evolutionary optimization algorithms [2] are used for automated optimization of geometries based on predefined constraints. These methods are computationally expensive and are suited only for improving aerodynamics, and cannot consider aesthetics. Evaluating aesthetics cannot be currently automated, and a human in the loop is required. Surrogate models enable fast aerodynamic evaluation and reduce the iteration time between aerodynamicists and stylists.

It is common practice in the automotive industry to use the Kriging model [3, 4] for predicting $c_d$, and ROM – via an approach called Proper Orthogonal Decomposition plus Interpolation (POD+I) [4] - for predicting fields (velocity, pressure & wall shear stress) in real-time. These two methods are also used in the present study as a reference for the results obtained via the neural networks and are referred to as the Kriging and ROM model in the remainder of this article. For both models, a few geometric features of a baseline geometry are parameterized. Geometries are sampled via Design of Experiments (DoE) and simulated via CFD. This data is then used to train the models. The trained models can be used for aerodynamic evaluation in the defined design space. The main disadvantage is that the models can work only for the parameters used during training. This study evaluates if deep learning models that work on arbitrary input geometries without explicit parameterization can be used for approximate aerodynamic evaluation, with error rates similar to those reported by Mrosek et al. [4]. The long-term goal would be to have models (e.g., one model per vehicle type) that can learn from multiple data sets created using different geometric parameterizations as well as from simulation results for vehicle shapes that have been evaluated during the development process. The resulting models increase in accuracy over time when more training data is gathered and allow for predicting the $c_d$ for arbitrary vehicle variations, thus yielding complete design freedom.

27/07/2021

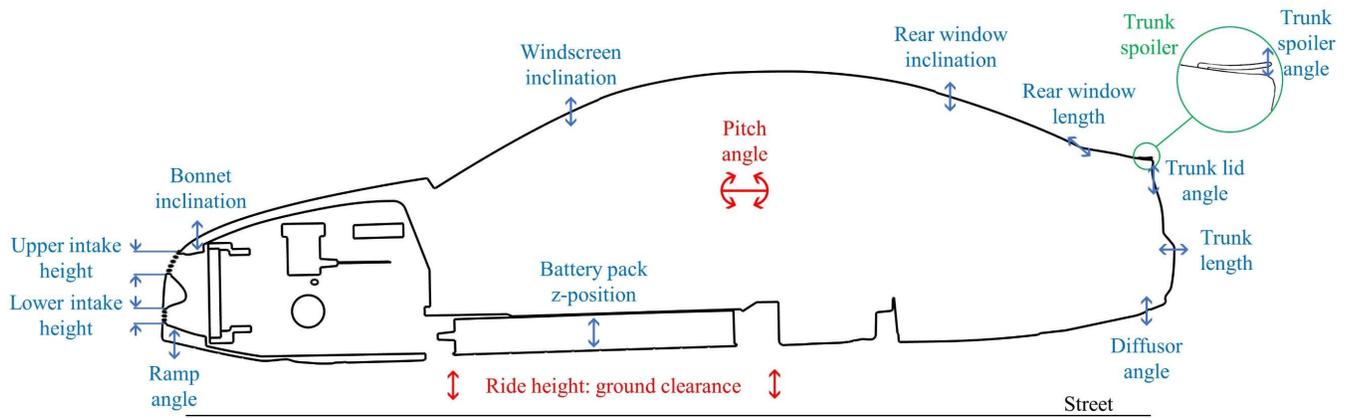

*Figure 1 Geometric parameters of the DrivAer data set. The parameters are color-coded: blue for local parameters, red for global parameters, and green for the Boolean parameter*

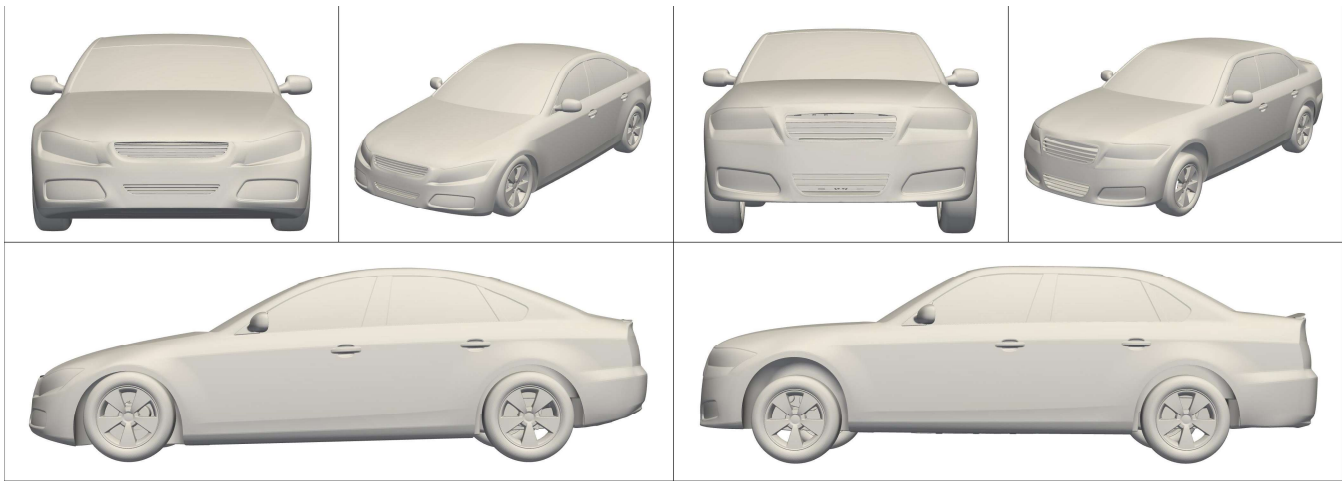

*Figure 2. Surface mesh corresponding to the parameters that lead to the minimum volume (left) and maximum volume (right) of the DrivAer geometry.*

Convolutional Neural Networks (CNNs) are commonly used to deal with spatial data and are good at extracting spatial features. However, simulation geometries are mostly made of unstructured meshes, and they cannot be directly represented in an orderly spatial grid, making it hard to use with CNNs. There have been multiple methods to make meshes amenable as input for neural networks; some of them are as follows: Baque et al. [5] remeshed the geometries to a poly-cube to construct a uniform simplified geometry and then uses geodesic convolutional neural networks to perform aerodynamic optimization. Thuerey et al. [6] used a binary mask to approximate Reynolds-averaged Navier–Stokes simulations of airfoils, where the points in a Cartesian grid are marked one if inside the geometry and zero if otherwise. SDF is used by Guo et al. [7] and Bhatnagar et al. [8] to represent the geometry for external aerodynamic evaluation. Qi et al. [9, 10] used discrete points to represent geometries for classification and segmentation tasks. There have also been recent methods capable of directly using manifold meshes [11], but they are memory intensive for larger meshes.

In this study, we evaluate if we can use SDF to represent geometries for aerodynamic evaluation. The data set is explained in the *DrivAer data set* section. In the *Methodology* section, we discuss the SDF representation, the network architecture, various training, and evaluation methods. In the *Results* section, we primarily focus on improving $c_d$ prediction, including experiments with predicting velocity fields in addition to predicting $c_d$. Additionally, we have performed an occlusion sensitivity study [12] to get an intuition on what the model learned. In summary, we evaluate if deep learning models can be used to predict $c_d$ while being agnostic to the solver used and the geometric representation.

## DrivAer data set

The DrivAer geometry was introduced by Heft et al. [13], such that there exists a public car geometry to simulate complex flows for performing aerodynamic research. The DrivAer model has fastback, notchback, and estate back variants, and we use a notchback variant with a simplified electric drive train introduced as part of the UPSCALE project [14]. Several geometric parameters that have a noticeable impact on the aerodynamics are chosen and are varied within a fixed range to produce the data set. We create a design of experiments by sampling parameters using Sobol sequences [15] from the constrained design space and split the samples as 70%, 15%, and 15% for training, validation, and test, respectively.

Figure 1 shows the 15 geometric parameters, and Table 1 shows the extents of their geometric variation. Mrosek et al. [16] have discussed the DrivAer data set used in detail. The trunk lid spoiler is a Boolean parameter, i.e., the trunk lid spoiler is either present or not. As the existing methods used within the automotive industry cannot deal with geometries created using multiple explicit parameterizations, we created two data sets, one for samples with spoiler and another for without spoiler. Figure 2 shows the minimum and maximum volume geometries based on the geometric parameters.

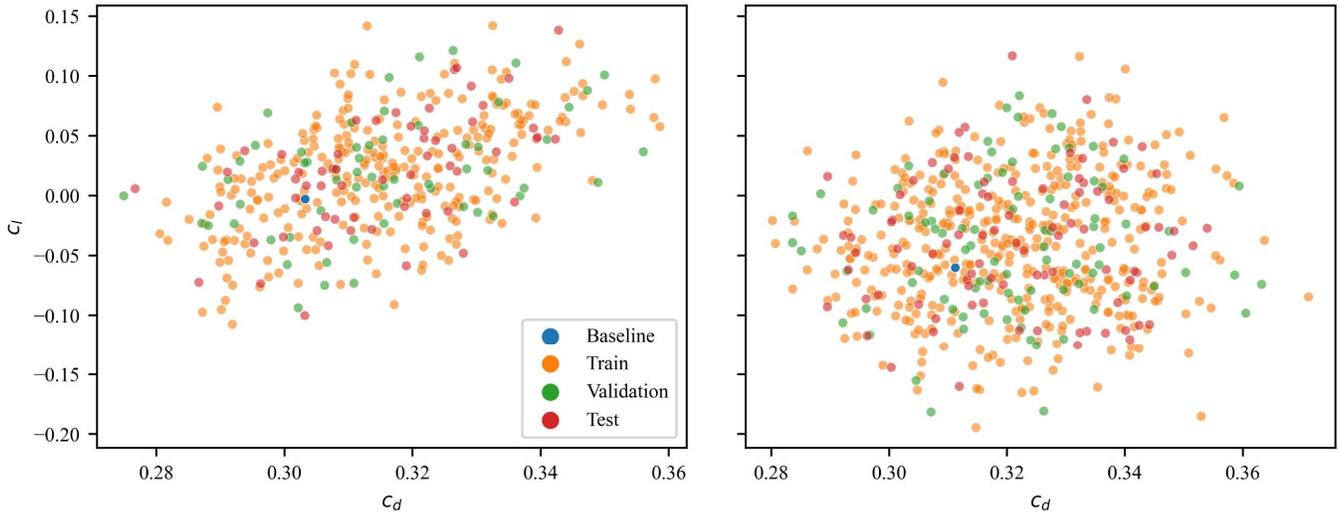

*Figure 3. Distribution of $c_d$ and $c_l$ for the different splits and the baseline geometry: DrivAer without spoiler data set (left) and DrivAer with spoiler data set (right).*

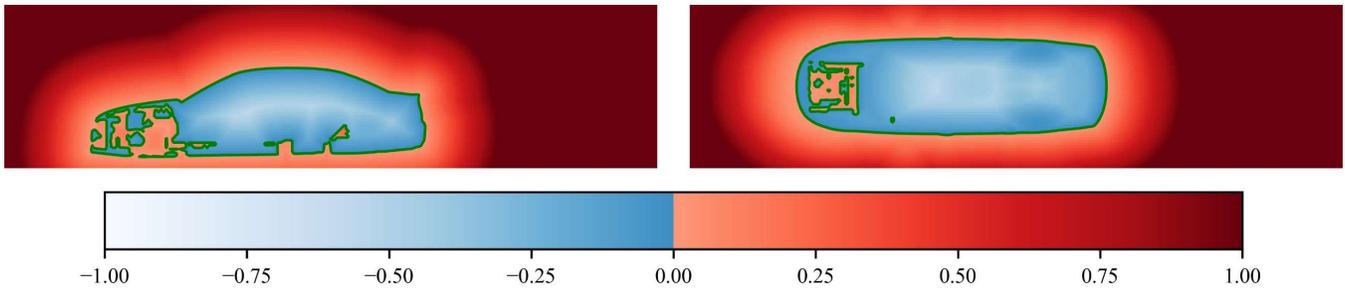

*Figure 4. SDF slices for the baseline DrivAer geometry. A green contour is present where the SDF is zero. The geometry is offset from the center because we generate SDF in the same volume where we are interested in making a velocity field prediction. We have clipped the SDF between -1 and +1 for easier visualization.*

The "DrivAer without spoiler" data set has 400 samples, and the "DrivAer with spoiler" data set has 600 samples. We performed Lattice Boltzmann simulations using Altair's ultraFluidX [17, 18], and each simulation took about 5.6 hours using eight NVIDIA V100 GPUs. We simulated four seconds of physical simulation time and averaged the aerodynamic coefficients & fields over the last two seconds. Figure 3 shows the distribution of $c_d$ and lift coefficient ($c_l$) for both data sets.

*Table 1. Summary of all the parameters of the DrivAer data set along with the minimum and maximum deviation values of the parameters with respect to the baseline. The trunk spoiler parameter is the only Boolean type parameter, and it controls if the spoiler exists. The ranges of the various parameters are chosen such that the values outside are either hard to manufacture or are not aesthetically pleasing.*

| Parameter name | Minimum value | Maximum value | Unit | Type |
|---|---|---|---|---|
| Ramp angle | -50 | 50 | mm | Continuous |
| Lower intake height | -30 | 30 | mm | Continuous |
| Upper intake height | -30 | 30 | mm | Continuous |
| Bonnet inclination | -50 | 50 | mm | Continuous |
| Windscreen inclination | -25 | 30 | mm | Continuous |
| Rear window inclination | -25 | 25 | mm | Continuous |
| Rear window length | -45 | 45 | mm | Continuous |
| Trunk spoiler | – | – | – | Boolean |
| Trunk spoiler angle | -8 | 16 | mm | Continuous |
| Trunk lid angle | -45 | 30 | mm | Continuous |
| Trunk length | -55 | 55 | mm | Continuous |
| Diffusor angle | -50 | 50 | mm | Continuous |
| Battery pack z-position | -30 | 30 | mm | Continuous |
| Ride height: ground clearance | -25 | 50 | mm | Continuous |
| Ride height: pitch angle | -1 | 1 | ° | Continuous |

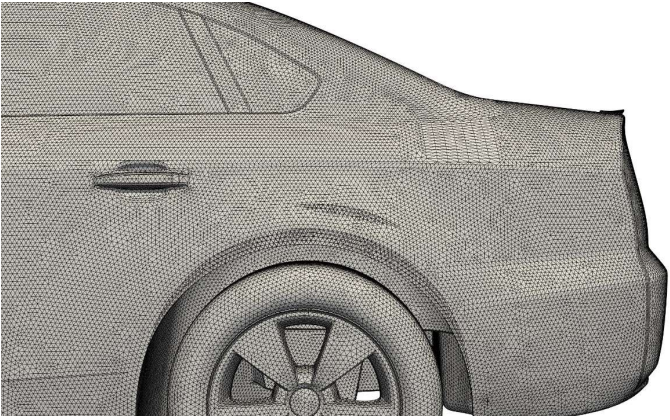

a) Original mesh

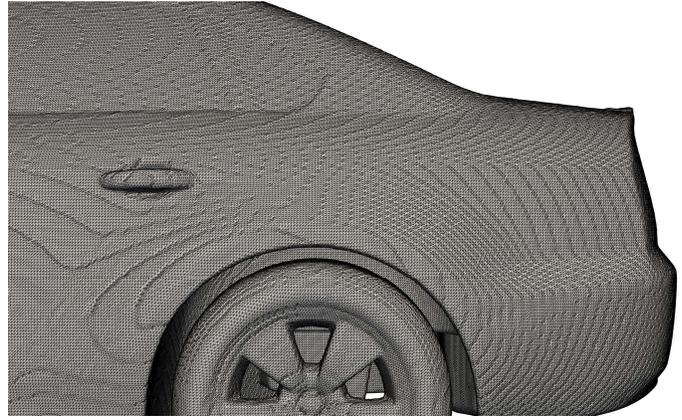

b) Manifold mesh

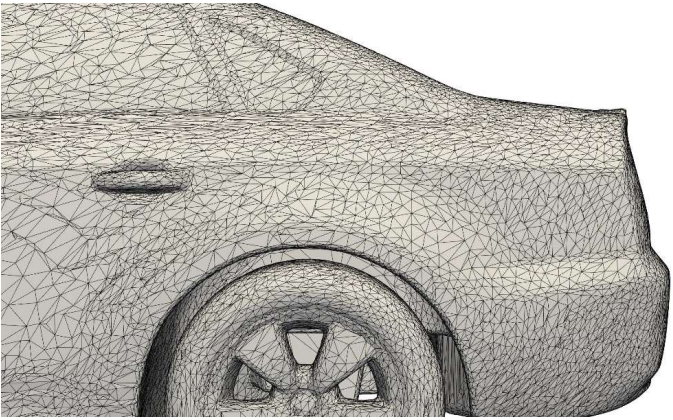

c) Simplified mesh

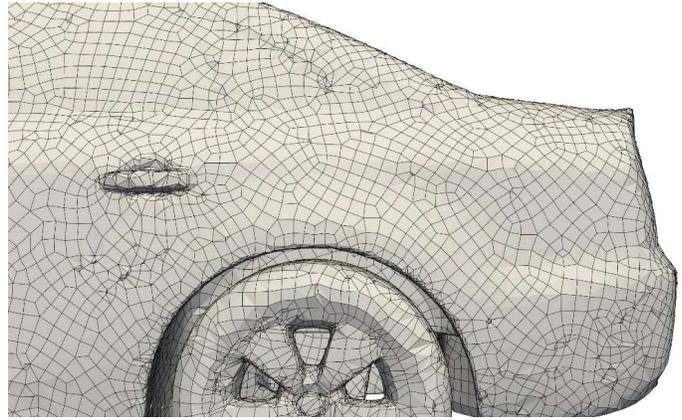

d) Remeshed mesh

*Figure 5. Examples of augmented meshes. The subfigures are arranged according to the order of the data augmentation process.*

## Methodology

### *SDF Representation*

SDF (Equation 1 and 2) is the distance between points in 3D space and the corresponding closest points on the geometry; the sign is positive for the points within the geometry and negative if otherwise.

$$sign(x) = \begin{cases} -1, & if\ x\ is\ inside\ the\ geometry \\ +1, & otherwise \end{cases} \quad (1)$$

$$SDF(x) = sign(x) . \min_{\forall p \in Z} |p - x| \quad (2)$$

Where:
$x \in \mathbb{R}^3$ = a cell center of the Cartesian grid
$Z$ = set of all points $z \in \mathbb{R}^3$ on the surface of the geometry

We generate SDF over a 3D Cartesian grid. From Figure 4, we see the 2D slice of SDF for a DrivAer sample. For the DrivAer data sets, the SDF can capture relevant geometric features and achieve reasonable error rates with a resolution of 256 x 64 x 64. Based on the scale of the geometric features we want to capture, we can adjust the resolution.

Watertight meshes are ideally suited to generate SDF. However, since our geometries are not watertight, we used an approximation method [19] to generate our SDF. Using this approximation method introduced some artifacts and noise, which are alleviated by the data augmentation introduced in the next section. It takes about an hour with a single CPU core to generate an SDF with a resolution of 256 x 64 x 64 for one geometry. It is possible to generate SDF near real-time by parallelizing SDF generation or by using other SDF generation libraries like OpenVDB [20].

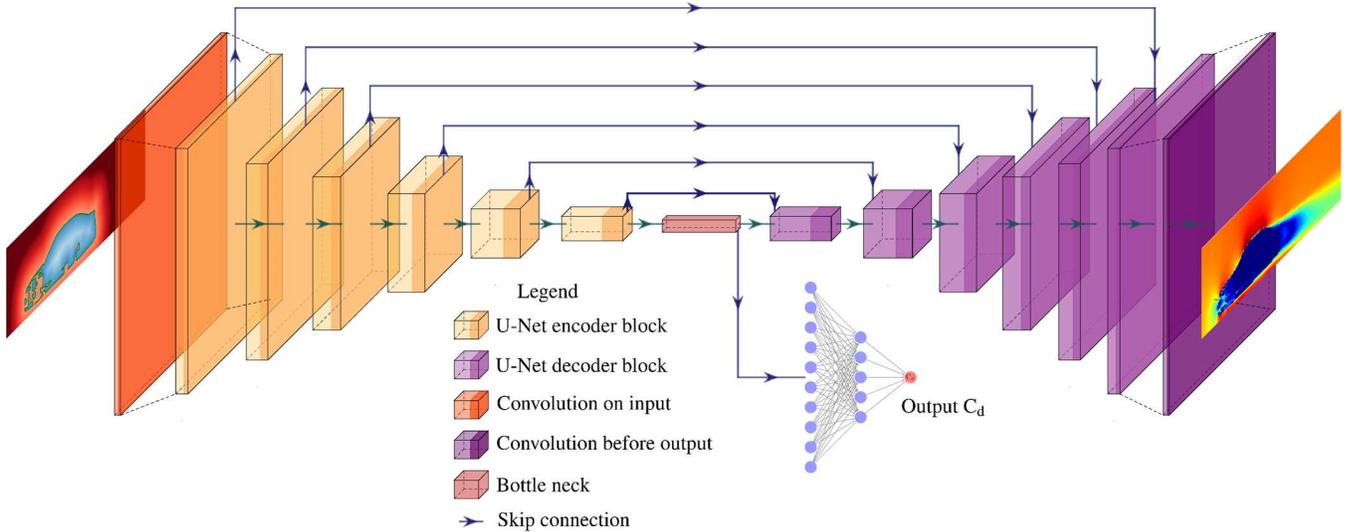

Figure 6. U-Net architecture illustration (using 2D slices through the 3D SDF input and 3D velocity field output). The architecture shows a multi-task learning model that can predict both the $c_d$ and velocity fields simultaneously. The velocity field decoder branch can be ignored when we predict only the $c_d$. In our models, we use three identical decoders for the three components of velocity. PlotNeuralNet [22] is used for the architecture illustration.

**Legend**
- U-Net encoder block
- U-Net decoder block
- Convolution on input
- Convolution before output
- Bottle neck
- Skip connection

Output $C_d$

### Data augmentation

Due to the enormous computational cost, it is time-consuming to generate a large amount of training data. We use data augmentation to increase the number of training samples at a low computation cost. Data augmentation also acts as a regularizer. It helps to prevent overfitting and to deal with the noise from SDF generation. Data augmentation is done by remeshing a geometry in multiple different ways to generate equivalent meshes. The augmented meshes have subtle differences compared to the original meshes; this forces the network to learn the geometric features that matter and deter the network from memorizing. Data augmentation also mimics the field implementation (i.e., when aerodynamicists or stylists use a trained model for aerodynamic evaluation), where the geometries presented to the network might use different mesh structures or mesh types. Since all our meshes used for the simulation are created in a very similar way, data augmentation makes the network more robust by exposing it to different mesh types.

Table 2. Summary of the number of samples in the DrivAer data sets.

|  | DrivAer without spoiler | DrivAer with spoiler |
|---|---|---|
| Total original samples | 400 | 600 |
| Training samples | 280 | 420 |
| Augmented training samples | 6,440 | 9,660 |
| Validation samples | 60 | 90 |
| Test samples | 60 | 90 |
| Total training samples with data augmentation | 6,720 | 10,080 |

It is computationally expensive to remesh the original meshes as they have about eight million cells. To reduce the computational cost, we simplify the original meshes before remeshing. We generate a manifold mesh [21] from the original mesh, which reduces the cell count to about five million. We simplify the manifold meshes to meshes having 250000, 450000, 500000, 550000, and 600000 faces.

The simplified meshes are remeshed seventeen times using different mesh types and meshing parameters. Figure 5 shows examples from different data augmentation steps, and Table 2 shows a summary of the number of samples in the data sets. We use meshes from all the steps for training.

### U-Net architecture

We use a modified U-Net architecture (Figure 6) to predict $c_d$ and the velocity fields concurrently. U-Net is an encoder-decoder architecture introduced by Ronneberger et al. [23] for medical image segmentation. There have been multiple variations and improvements for various applications [24, 25], including CFD [6]. The encoder part of a U-Net has multiple blocks that progressively down sample and extract a growing number of feature maps with increasing complexity. The encoder extracts all the relevant information from the input SDF and outputs a global feature vector. Then, the global feature vector is passed through decoders to predict the $c_d$ and the velocity fields. In essence, the global feature vector can predict any other parameter or field when trained with a suitable decoder.

The velocity field decoder is a mirror of the encoder part. It progressively up samples the feature map size and reduces the number of feature maps to predict the velocity field. The decoder blocks have skip connections from the corresponding encoder blocks. The skip connections enable the decoder to reuse the feature maps learned by the encoder. The $c_d$ decoder flattens the global feature vector, then passes through a dropout layer, a fully connected layer, ReLU activation, a dropout layer, a batch normalization layer, and an output fully connected layer of one neuron.

A sequence of ReLU activation, squeeze & excitation block [26], convolutional layer with dilations, max-pooling layer (with a stride of two), and batch normalization layer is a single U-Net encoder block. The convolutional and the max-pooling layer of the U-Net encoder block are replaced by a transpose convolutional layer (with a stride of two) for the velocity field decoder block. We have three separate velocity field decoders for the three components of velocity. The encoder has a convolutional layer followed by six U-Net encoder blocks.

Typically, not all the feature maps and all the spatial information is relevant for a specific geometry. Squeeze & excitation blocks explicitly model selective weighting of most relevant information, akin to self-attention methods that learn to focus on the most relevant information selectively. We use concurrent spatial and channel squeeze & excitation block [27], a variation of squeeze & excitation block, which weighs the importance of spatial regions and the importance of feature maps concurrently.

When predicting both velocity fields and $c_d$, we have 100 million trainable parameters. We do not use the velocity field decoders when predicting only the $c_d$; in this case, we have about 23 million trainable parameters. We use either Adam or NAdam optimizer and "ReduceLROnPlateau" learning rate decay implementation from TensorFlow [28]. ReduceLROnPlateau reduces the learning rate by a factor when there is no improvement in the validation loss for a predefined number of epochs. We perform early stopping based on the $c_d$ validation loss. We later restore the weights from the best performing epoch.

### Loss and error metrics

The loss function used is the Weighted Mean Squared Error (WMSE - Equation 3 and 4). When we compare augmented meshes with original meshes, we can see slight variations. To account for this error, we weigh the augmented meshes lower than the original meshes, which is akin to informing the model that we are less confident of the augmented meshes compared to the original meshes. We also weigh multiple tasks differently. When predicting both $c_d$ and velocity fields, we weigh the loss for velocity field prediction lower because we prioritize $c_d$ prediction.

In the DrivAer data set, only a small region of the total velocity field varies due to the geometric variations, and the velocity field in most regions is relatively constant. The network can get a good portion of accuracy by predicting the constant regions, but the network cannot learn much from these regions. We weigh the velocity field such that it forces the model to give higher importance to the regions in the velocity field that have a higher variance across the training samples. The weight for velocity field loss is calculated by computing the variance of the velocity field (cell-wise) in the training set and then scaling it between 0.5 and 4. This weighting forces the network to focus more on the vehicle's wake and less on regions inside the geometry and the regions with low variance.

We evaluate the models by comparing the $R^2$ score (also called the coefficient of determination, Equation 5) on the test set for $c_d$

prediction, with the best possible $R^2$ score being one. We use the Mean Absolute Error (MAE) as in Equation 6 for a straightforward interpretation of the $c_d$ error. For $c_d$ estimation in wind tunnel experiments, MAE less than 0.005 with respect to wind tunnel measurements is considered acceptable [29]. To find the worst $c_d$ estimation, we also compute the Maximum Absolute Error (Max AE). For evaluating the predicted velocity fields, we use the relative L2-error (Equation 7).

$$\text{WMSE}_{c_d} = \frac{1}{n_{test}} \sum_{i=1}^{n_{test}} w^i \left( c_d^i - \widehat{c_d}^i \right)^2 \tag{3}$$

$$\text{WMSE}_u = \frac{1}{n_{test}} \frac{1}{n} \sum_{i=1}^{n_{test}} \sum_{j=1}^{n} w_j^i \left( u_j^i - \hat{u}_j^i \right)^2 \tag{4}$$

$$R^2 = 1 - \frac{\sum_{i=1}^{n_{test}} (c_d^i - \widehat{c_d}^i)^2}{\sum_{i=1}^{n_{test}} (c_d^i - \overline{c_d})^2} \tag{5}$$

$$\text{MAE} = \frac{1}{n_{test}} \sum_{i=1}^{n_{test}} \left| c_d^i - \widehat{c_d}^i \right| \tag{6}$$

$$L2 = \frac{1}{n_{test}} \sum_{i=1}^{n_{test}} \frac{\| u_j^i - \hat{u}_j^i \|_2}{\| u_j^i \|_2} \tag{7}$$

Where:
$n_{test}$ = Number of test samples
n = Number of data points: Number of cells times three (number of velocity field components)
$c_d^i$ = True drag coefficient value
$\overline{c_d}$ = Mean true drag coefficient value
$\widehat{c_d}^i$ = Predicted drag coefficient value
$u_j^i$ = True velocity field vector
$\hat{u}_j^i$ = Predicted velocity field vector
$w^i$ = Sample/data point weight

### Occlusion Sensitivity

When CNNs first gained mass exposure, it was regarded as a black box, and many were hesitant to use it. Methods like Grad-CAM [30] and guided backpropagation [31] were later developed to understand what and how the network is learning. Visualizing the filters learned by the network and seeing where specific filters produce high activation values is a straightforward way to do this. However, this method is not of much use in our case because SDF carries both global and local information, and it is also hard to visualize & understand 3D filters.

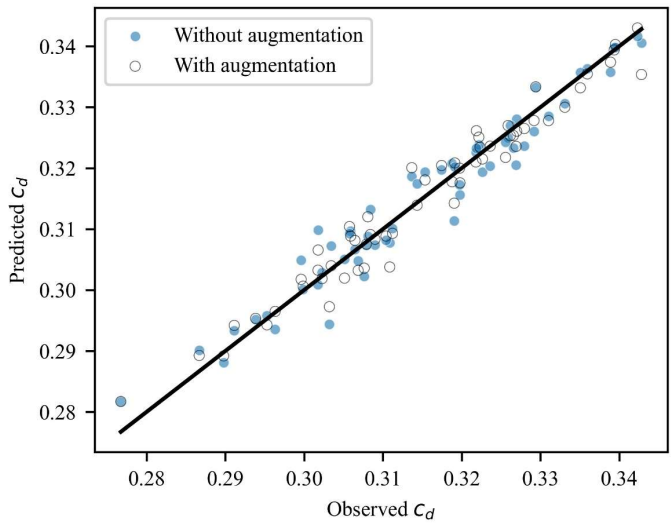 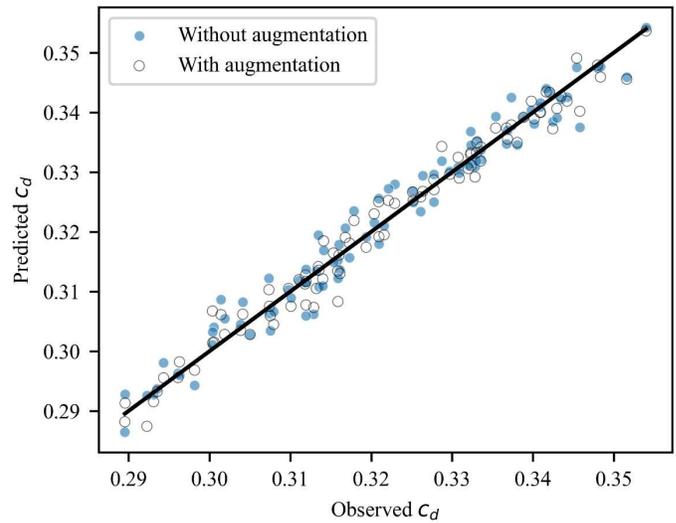

a) Model trained using DrivAer without spoiler data set and evaluated for DrivAer without spoiler data set

b) Model trained using DrivAer with spoiler data set and evaluated for DrivAer with spoiler data set

*Figure 7. Correlation plot (for test samples) between the true $c_d$ and predicted $c_d$. We compare the best model trained with data augmentation against the best model trained without data augmentation.*

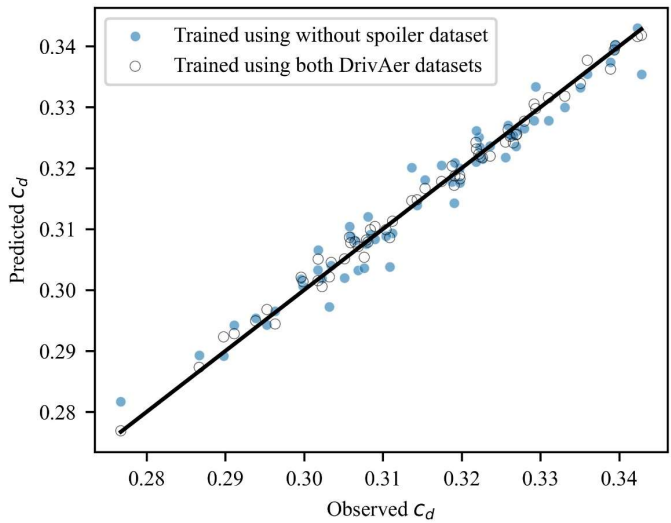 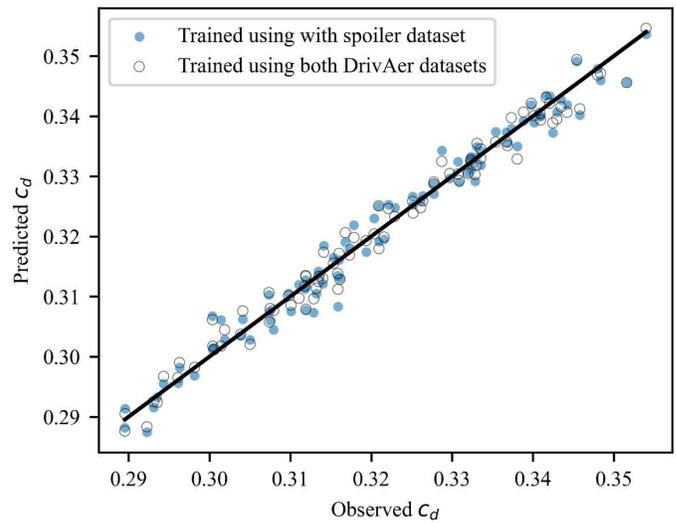

a) Model evaluated for DrivAer without spoiler data set

b) Model evaluated for DrivAer with spoiler data set

*Figure 8. Correlation plot (for test samples) between the true $c_d$ and predicted $c_d$. We compare the best model trained using both the DrivAer data sets against the best model trained only using the data set it is evaluated on. All the models were trained using data augmentation.*

Occlusion sensitivity [12] is a method that can be used for evaluating which regions of the input SDF are crucial for the model to make an accurate prediction. By methodically occluding certain areas of the input SDF, we can determine the regions that lead to an increase in the error. First, we set the values of all the elements in a cube of edge length ten to zero in the input SDF, and then we move the cube around the SDF with a stride of five and evaluate the MAE on the test set. Finally, we scale the MAE between zero and one for a more straightforward interpretation. The regions with values close to one are crucial for the model to make accurate predictions, and regions with values lower than 0.1 have a negligible impact on the predictions.

## Results

### DrivAer individually trained model results

We trained models using either only DrivAer with spoiler data set or DrivAer without spoiler data set and evaluated only on the test samples of the corresponding data set. We also evaluated if the model improved with data augmentation. From Table 4, we can see that the model trained using DrivAer with spoiler data set with data augmentation has its metrics very similar to the Kriging model. All other models performed worse when compared to the Kriging model. From the correlation plots (Figure 7) and the metrics, we can see that models trained using data augmentation are better (about 13% and 19% concerning MAE) than models trained without augmentation.

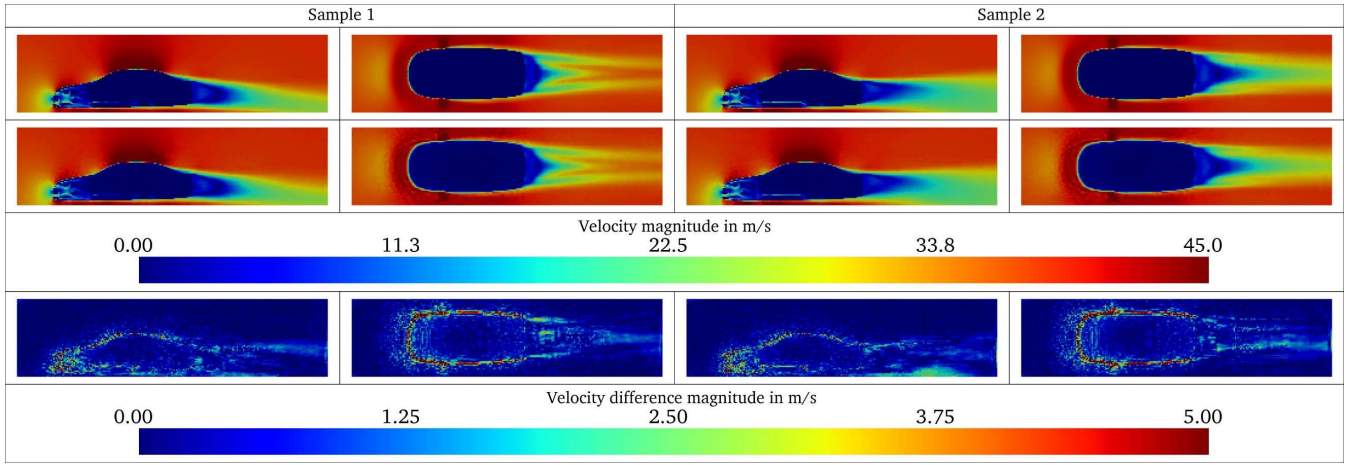

*Figure 9. Comparing slices of CFD results with velocity field prediction from multi-task learning model trained on DrivAer with spoiler data set. The first row is the true CFD results, the second row is the prediction, and the third row is the absolute error. We have chosen two samples that have different velocity profiles.*

Although the best models trained here do not use squeeze & excitation blocks, the models using squeeze & excitation blocks had very close results. As models trained using data augmentation are better than models trained without data augmentation, we will only discuss models trained with data augmentation in further experiments.

*Table 3. Evaluation of test metrics and comparison with the Kriging model for $c_d$ prediction. Models trained without data augmentation were trained using one NVIDIA P100 GPU. Models trained with data augmentation were trained using three NVIDIA P100 GPUs.*

| Model | Evaluation data set | $R^2$ score | MAE | Max AE | Training time [h] |
|---|---|---|---|---|---|
| Our model without augmentation | Without spoiler | 0.951 | 0.0025 | 0.0088 | 1 |
| | With spoiler | 0.967 | 0.0023 | 0.0083 | 3 |
| Our model with augmentation | Without spoiler | 0.961 | 0.0022 | 0.0074 | 9 |
| | With spoiler | 0.975 | 0.0019 | 0.0076 | 17 |
| | Without spoiler | 0.990 | 0.0013 | 0.0026 | 31 |
| | With spoiler | 0.980 | 0.0017 | 0.0060 | |
| Kriging | Without spoiler | 0.968 | 0.0020 | 0.0070 | - |
| | With spoiler | 0.977 | 0.0019 | 0.0080 | - |
| Legend | | | | | |
| Trained using DrivAer without spoiler data set | | | | | |
| Trained using DrivAer with spoiler data set | | | | | |
| Trained using both DrivAer with spoiler and DrivAer without spoiler data set | | | | | |

### Model trained using both DrivAer data sets

We trained the models using both DrivAer data sets to see if the models improve by leveraging information from both data sets simultaneously. It turned out that the models improved significantly compared to the Kriging model and the models trained using only one DrivAer data set, as seen from the metrics in Table 4 and the correlation plot (Figure 8). We can see an improvement of about 35% for DrivAer without spoiler data set and 11% for DrivAer with spoiler data set over the Kriging model. The inclusion of squeeze & excitation blocks is the only change in the architecture.

It is likely that when we use both the DrivAer data sets, the results improve because there are more training samples, and in our case, all but one geometric feature are common between the data sets. In contrast, the conventional Kriging model has one primary drawback: we can train only with a single data set. This experiment strongly indicates that our model can make more accurate predictions when trained using similar data sets by reusing and refining learned feature maps.

### Multi-task learning model

*Table 5. $c_d$ prediction for multi-task learning models trained on DrivAer with spoiler data set - evaluation and comparison with existing models. L2-error is used for evaluating the fields and the other metrics for evaluating $c_d$. The models were trained using four NVIDIA P100 GPUs.*

| Model | $R^2$ score | MAE | Max AE | L2-error [%] | Training time |
|---|---|---|---|---|---|
| Our model | 0.969 | 0.0023 | 0.0068 | 2.80 | 27 hours |
| Kriging | 0.977 | 0.0019 | 0.0080 | - | - |
| ROM | - | - | - | 1.59 | - |

We trained a multi-task learning model that can concurrently predict $c_d$ and velocity fields using the DrivAer with spoiler data set. From Table 5, we can see that the model performed worse than the previous models. Nevertheless, we can see that the model can predict both the velocity fields (with a 1.21% absolute increase in L2-error as compared to ROM) and $c_d$ (with a 21% relative increase in MAE compared to the Kriging model) simultaneously. Figure 9 shows the velocity field prediction for two samples from the test set. From the figure, we can see that the model can capture the large-scale features for both test samples. We can see artifacts predominantly close to the geometry surface, which could be the checkerboard artifacts [32] caused by transpose convolutional layers. One of the difficulties in training multi-task learning models is that the tasks are of varying complexity and require different numbers of training epochs. To reduce these conflicts, we weighed the velocity field loss lower, weighed the velocity field based on variance, and applied learning rate decay based only on the $c_d$ validation loss.

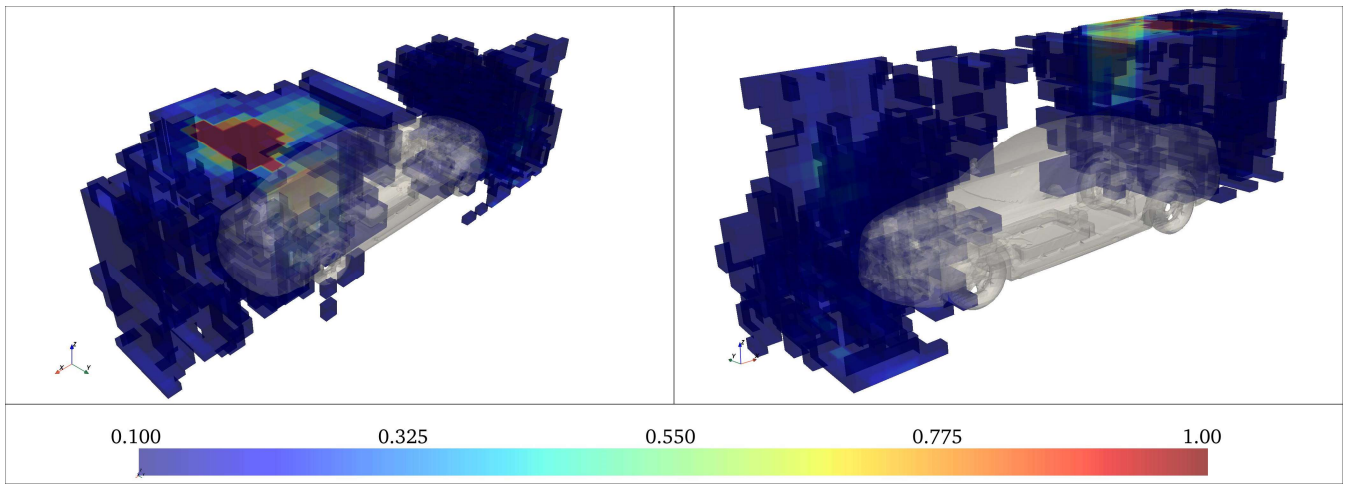

*Figure 10. Visualization of occlusion sensitivity maps from two views for a model trained using both DrivAer data sets. The occlusion map is thresholded from 0.1 to 1.0. The regions with a value lower than 0.1 have little to no effect on the prediction. The thresholded region corresponds to about 22% of the total volume. It is concentrated mainly around the front and rear of the car, which corresponds to the regions with most of the parameterized features. There is a break in the middle of the car where there are only a few geometric features.*

Multi-task learning models improved only for about 100 to 120 epochs compared to 300 to 500 epochs for models trained solely for predicting $c_d$. We suspect that this is either because (1) the network learns relatively faster with the velocity fields or (2) that after the early epochs, the two tasks are no longer complementing one another and are unable to improve. Multi-task learning shows promise, but further research is needed to find better training methods and approaches to balance the tasks. Improvement in predicting the velocity field is among the future works.

## Occlusion sensitivity

We performed an occlusion sensitivity study on our best models. For an easier understanding of the 3D occlusion map, we have visualized after clipping the occlusion map values between a few specific ranges. From Figure 10, we can see a concentration of regions of importance in the front and the rear of the car, and about 22% of cells have values higher than 0.1. In addition, we can see a distinct gap in the middle which shows that the model has discovered that most of the geometric features are in the front or rear of the geometry.

We can see from Figure 11, most of the regions with a high impact on the $c_d$ are close to the domain boundary, and surprisingly at first sight, the network does not prefer the regions near the surface. However, we need to keep in mind that SDF contains both local & global information. The cells with the highest importance values have global information from the rear of the geometry. We can also infer from Figure 11 that the geometric features in the rear affect the $c_d$ prediction more than the other geometric features. The model can use the regions with values greater than 0.9 to estimate eight of the fourteen geometric features.

It is not an anomaly that the best model prefers the regions close to the domain boundaries as this behavior was consistent among multiple well-performing models. The network can estimate all the relevant geometric features from the global information present in a few fixed regions in space. This workaround of exploiting the global information was probably more straightforward for the model to learn than learning to estimate the geometric features by following the vehicle surface. Guo et al. [7] showed that models trained using SDF could perform better than binary mask representations for the same model complexity. A reason for this might be that the model does not have to learn to follow the geometries' surface when using SDF to represent geometries.

It is well known that models sometimes learn workarounds, wrong features, or noise but still make reasonable predictions. It is also common to use multiple regularization methods to prevent this. For our use case, the model using this workaround might not be an issue when we train models on data sets for one vehicle geometry. However, using the workaround might be of concern when we expect the trained model to transfer learned geometric features from one vehicle geometry to another, as we may have unintentionally introduced a bias regarding the orientation and scales of the geometric features.

The long-term goal would be to have a vast repository of simulations from multiple cars (of the same vehicle type) and the ability to predict approximate aerodynamic results for new vehicle shapes. This goal can be achieved only if the model can accurately transfer geometric features. Towards this aim, we can do transformations like rotation, scaling, translation, etc., on the input geometry. This can pressure the network to learn to recognize the same geometric feature from different orientations. However, this data augmentation method is rather impractical because simulating multiple augmented geometries would be computationally expensive.

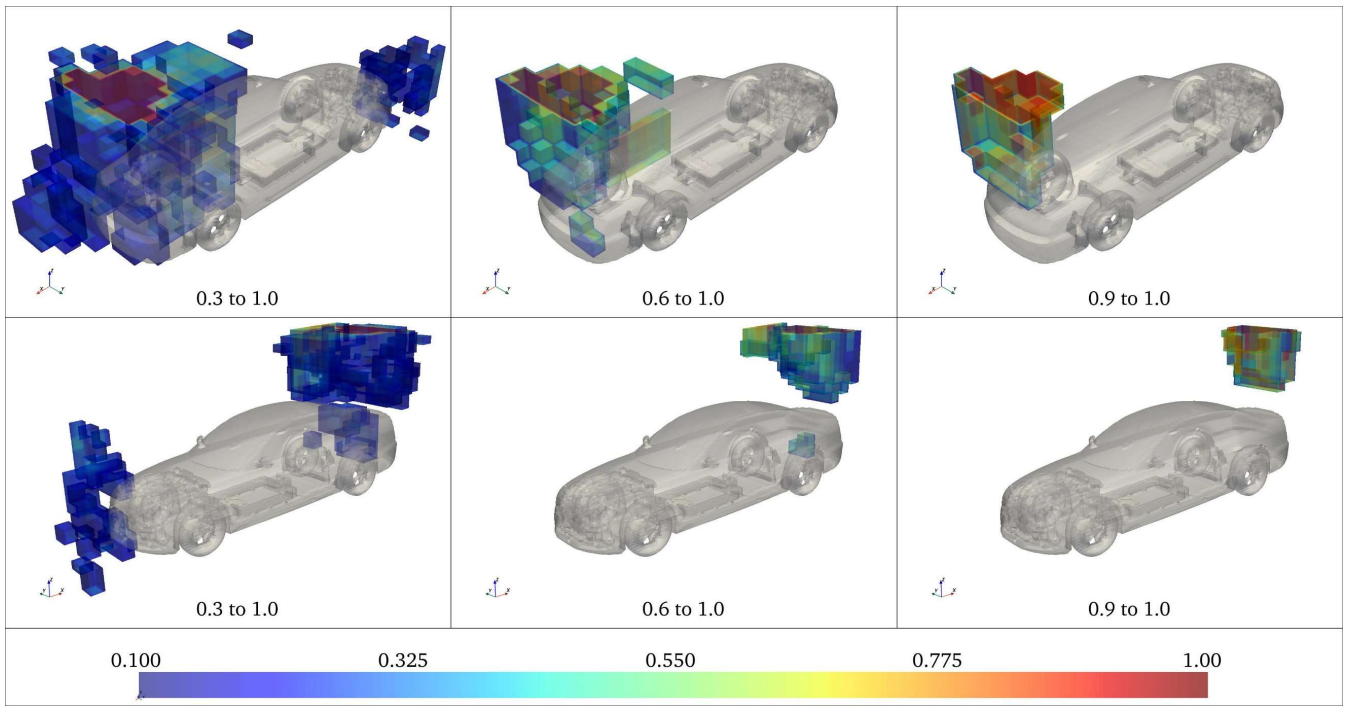

*Figure 11. Same as Fig. 10. The volume is thresholded in three different ranges (see captions) for easier understanding. From the thresholded volumes, we can infer that the geometric features in the rear of the car affect the $c_d$ more than the geometric features in the front of the car.*

As a more practical approach that does not require additional simulations, we can make the model robust to affine transformations using an autoencoder setup. We could perform numerous transformations on the car geometries and generate SDF at a low computation cost. The augmented SDF could be used to train an autoencoder, and we could predict the $c_d$ and velocity fields using the latent representation. This method could result in a model that might be more robust to transformations and be better at transferring geometric features. If we doubt that the model is not learning the intended geometric features properly, this will be an area to investigate in the future.

## Conclusions

In summary, we demonstrated that neural networks could accurately predict the drag coefficient $c_d$ and velocity fields of an arbitrary input mesh represented as an SDF without explicit parameterization. Our best models outperformed the Kriging model in predicting $c_d$ for both data sets with at least an 11% MAE improvement. Although models trained without data augmentation made reasonable predictions, we demonstrated that data augmentation improves results and that we can train deep learning models without a large number of training simulations.

Our models trained on individual DrivAer data sets performed worse or similar to the Kriging model, but our models trained using both the DrivAer data sets outperformed the Kriging model significantly. We demonstrated that the network improves with additional samples from a similar data set, which is a notable advantage over the Kriging model.

We demonstrated that velocity fields could be predicted concurrently with $c_d$ and presented a method of weighting the velocity field based on its variance across training samples, which forces the network to focus on vital regions like the wake of the car. Multi-task learning shows promise, even though we were unable to produce results better than the existing models. Better training methodology and strategies to balance both tasks are some of the areas in need of improvement.

We performed an occlusion sensitivity study to get an insight into how the network is learning. We showed that the model has identified regions where the geometric parameters are present, and the model prioritized the geometric parameters in the DrivAer's rear. We also realized that the model had learned a workaround for estimating the geometric parameters by just looking at a few fixed regions close to the domain boundary, thus exploiting that the SDF carries both global and local information. Using this workaround may not be a concern when working with datasets created from a single-vehicle geometry. However, it might be a problem when we want the model to transfer geometric features between multiple geometries.

The DrivAer data set was generated using explicit parameterization and was used for this study as, currently, no real-world data set generated without explicit parameterization exists. Such a data set could be, for example, a historical data set containing all simulation results for one particular vehicle model gathered during the exterior shape development, potentially even across various generations of that particular vehicle model. Future work foresees the creation of such a data set generated without explicit parameterization. Other future works include improving multi-task learning and experimentation with larger data sets.

## Contact Information

**Markus Mrosek**
Volkswagen AG
Group Innovation
Wolfsburg
markus.mrosek@volkswagen.de
Tel. +49 5361 9-44407

## Disclaimer